\documentclass[acmsmall,screen]{acmart}
\usepackage{xcolor}
\usepackage{xltabular}

\AtBeginDocument{%
  \providecommand\BibTeX{{%
    \normalfont B\kern-0.5em{\scshape i\kern-0.25em b}\kern-0.8em\TeX}}}

\setcopyright{acmcopyright}

\acmJournal{TOCHI}

\begin{document}

\title{GAVIN: Gaze-Assisted Voice-Based Implicit Note-taking}

\author{Anam Ahmad Khan}
\email{anamk@student.unimelb.edu.au}
\orcid{0000-0002-1620-1902}
\author{Joshua Newn}
\email{joshua.newn@unimelb.edu.au}
\author{Ryan M. Kelly}
\email{ryan.kelly@unimelb.edu.au}
\author{Namrata Srivastava}
\email{srivastavan@student.unimelb.edu.au}
\author{James Bailey}
\email{baileyj@unimelb.edu.au}
\author{Eduardo Velloso}
\email{eduardo.velloso@unimelb.edu.au}
\affiliation{%
  \department{Computing and Information Systems}
    \institution{The University of Melbourne}
  \city{Melbourne}
  \country{Australia}} 
  
\renewcommand{\shortauthors}{Khan, et al.}

\begin{abstract}

Annotation is an effective reading strategy people often undertake while interacting with digital text. It involves highlighting pieces of text and making notes about them. Annotating while reading in a desktop environment is considered trivial but, in a mobile setting where people read while hand-holding devices, the task of highlighting and typing notes on a mobile display is challenging. In this paper, we introduce GAVIN, a gaze-assisted voice note-taking application, which enables readers to seamlessly take voice notes on digital documents by implicitly anchoring them to text passages. We first conducted a contextual enquiry focusing on participants’ note-taking practices on digital documents. Using these findings, we propose a method which leverages eye-tracking and machine learning techniques to annotate voice notes with reference text passages. To evaluate our approach, we recruited 32 participants performing voice note-taking. Following, we trained a classifier on the data collected to predict text passage where participants made voice notes. Lastly, we employed the classifier to built GAVIN and conducted a user study to demonstrate the feasibility of the system. This research demonstrates the feasibility of using gaze as a resource for implicit anchoring of voice notes, enabling the design of systems that allow users to record voice notes with minimal effort and high accuracy.

\end{abstract}


\maketitle

\section{Introduction}

With the wide popularity of e-reader devices and the increasing availability of popular titles in e-book form, digital reading has become a staple of modern life. Though e-book readers and tablets make it easier than ever to consume information from a virtually infinite library of titles, creating textual annotations---one of the most common tasks that readers do when interacting with text~\cite{ben2014,Cheema2016AnnotateVisC, porterodonnell2004}---is not as easy. Whereas note-taking on desktop devices is trivial, note-taking on mobile devices such as tablets faces many unique challenges, particularly when the user is away from a desk or without a stylus. Typically, the user must carefully touch down on the first word to be highlighted, drag to select the rest of the passage, and then use a virtual keyboard to type the note. Not only does this take the reader away from the text in order to input their thoughts, the different contexts of use and ways in which users hold the device may further hinder this task.

Given this challenge, voice has emerged as a potential input modality to address many of these issues. Because it is hands-free, not only it is suitable for interaction with mobile devices, users can also make the annotation without taking their attention away from the text. Though it is not suitable for every context (e.g.~classrooms, quiet public spaces), voice offers many additional opportunities for annotating digital text. First, compared to typing, voice notes have been shown to trigger generative processes that leads the reader to discuss more content and to elaborate more deeply on the content by relating it to prior life experiences \cite{khan2020}. This change in the reader's behaviour due to the switch in input modality leads to a better conceptual understanding of the digital text \cite{khan2020} and assists readers in the mental encoding and retrieval of the content---what is known as the ``production effect''~\cite{Berry1983,Cowan2017,Forrin2017,MacLeod2011}. Second, the advent of voice-based personal assistants has made interacting using voice more commonplace and socially acceptable \cite{zhou2012,wulf2014}. Third, advances in natural language processing have made translating between speech and text faster and more accurate, with new techniques even being able to extract additional information from audio, such as affect~\cite{schuller2018speechaffect}. Lastly, speaking has been found to consume fewer cognitive resources than writing, freeing the reader's working memory for other tasks~\cite{Grabowski2005}. 

However, the use of voice for text annotation also introduces new challenges. In general, applications that support this feature suffer from two critical challenges---anchoring and retrieval. \textit{Anchoring} relates to the mapping between a voice note and the region of content to which it refers. In written notes, the user typically highlights a text region and types their note, but if they are required to select this region manually when taking voice notes, the reading flow is broken. \textit{Retrieval} refers to how users revisit their notes for review. Whereas it is trivial to skim text notes to find the one the user is after, it may be harder to re-find voice notes, particularly if they need to be listened to in sequential order to find the desired recording.

\begin{figure}[tbp]
  \centering
  \includegraphics[width=0.85\columnwidth]{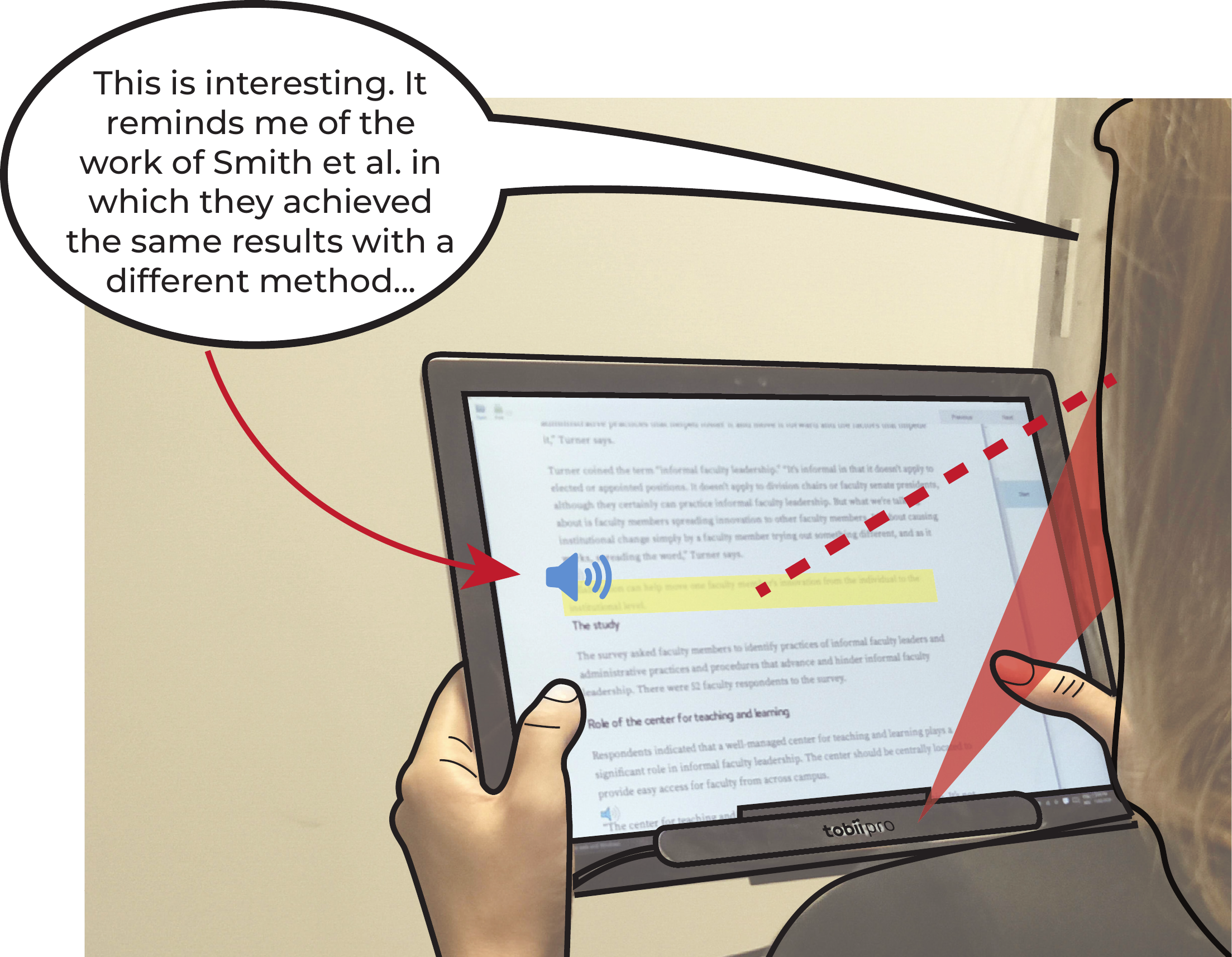}
  \caption{In GAVIN, users make verbal comments about what they read and the system uses a machine learning model to infer to which passage the comment referred to, automatically attaching the voice note to it.}
  \label{fig:gavin}
\end{figure}

In this paper, we provide a solution for making the task of voice annotation more implicit, making it easy and convenient for readers to interact with digital text on mobile devices. We propose to leverage the reader’s gaze as a resource for anchoring voice notes to text passages. The use of gaze is a natural fit considering that it is reflective of the user's visual attention \cite{Sibert2000, vo2010} and has long been used to reveal the reading strategies and comprehension ability of readers \cite{Cheng2015, kunze2013, liao2017, Buscher2008}. Hence, we present the design, implementation, and evaluation of GAVIN, a \textbf{G}aze-\textbf{A}ssisted \textbf{V}oice-based \textbf{I}mplicit \textbf{N}otetaking application that enables users to seamlessly anchor voice notes to text passages in two steps. First, the reader records a note using voice as the input modality. Second, the system implicitly maps the recorded voice note to the reference text passage by leveraging a machine learning model that processes the reader's gaze data (Figure ~\ref{fig:gavin}). The reader can then retrieve the note directly from the corresponding region. Our motivation is that implicit annotation on digital documents diminishes the need to engage in manual effort to anchor voice notes to text passage, allowing the readers to maintain focus on the reading material.

We ground the design of GAVIN on a qualitative user study with 12 participants in which we aimed to understand key issues that arise in note-taking on digital documents. We found that notes range from short comments to comprehensive reflections, but in all cases, it is crucial to be able to anchor them to text passages and to be able to retrieve them easily at a later stage. Further, we contribute a data processing pipeline that can map voice notes to text passages implicitly, without any additional input from the reader. For the scope of this paper, a text passage is defined as a paragraph component of a page. We hypothesised that we could predict the text passages to which a voice note refers by analysing the readers' gaze data as they spoke. To explore this possibility, we collected a dataset with 32 participants in which they recorded voice notes as they read an article of interest. We observed different types of gaze behaviour during voice note-taking, which meant that anchoring voice notes to text passages is not as straightforward as merely selecting the one where the reader was looking when speaking. To overcome this challenge, we engineered spatially based gaze and temporal features and built a Random Forest classifier based on our collected dataset. With our proposed approach, we were able to anchor voice notes to the right passage with an $AUC$ score of \textit{0.89}. To demonstrate the performance of our approach, we further implemented two baseline approaches to implicitly annotate voice notes. The first baseline approach leverages the position data (vertical scrollbar value and page number) of the document and annotates the recorded voice note to the top text passage of the page being read. The second baseline approach leverages the gaze trace to anchor the voice note to the text passage the reader has looked at the most while recording the note. We observed that our proposed approach of a classifier trained on spatially based gaze and temporal features outperformed the baseline approaches by an average $AUC$ score of \textit{0.25}. Further, we used our proposed classifier to build GAVIN, an interactive gaze-assisted voice note-taking application that implicitly anchors the recorded voice notes to text passages. We demonstrate the feasibility of GAVIN in a user study with 12 additional participants, in which we collected user feedback about potential opportunities for future use and barriers to adoption. 

In summary, we provide the following contributions through this paper:

\begin{enumerate}
     \item A classifier trained on spatially based gaze and temporal features engineered from the data of 32 participants while reading text that matches vocal utterances to text passages.
     \item  A comparison of our approach to two baseline methods based on scrolling behaviour and the gaze trace of the user to implicitly map voice notes to text passages.
    \item A working prototype gaze-assisted voice note-taking system that implicitly anchors voice notes automatically to the text, enabling easier retrieval.
\end{enumerate}

\section{Related Work}
Our research draws on prior work related to voice note-taking systems, gaze- and voice-based interactions, implicit document annotation, and the use of gaze data to build machine learning models.

\subsection{Voice Note-Taking Systems}

A considerable amount of work has been done on building applications that support annotation of digital documents in both textual and audio format. Note-taking applications like \textit{Dynomite}~\cite{Wilcox1997}, \textit{NiCEBook}~\cite{Brandl2010} and \textit{Evernote}\footnote{\url{https://evernote.com/}} facilitate the structuring, organising and sharing of textual annotations. In recent years, with the wide acceptance of voice as an input modality and due to its hands-free nature, voice has emerged as an alternative modality for recording notes to annotate digital documents. Voice annotation offers some advantages over textual annotation. First, annotation using voice as an input can assist readers in gaining a higher conceptual understanding of the digital text \cite{khan2020}. A recent study conducted by Khan et al.~\cite{khan2020} investigated the effect of the modality used for taking notes on readers' conceptual understanding of the text. The study compared two commonly used input modalities---\textit{keyboard} and \textit{voice}. The results of the study suggest that compared to typing notes using the keyboard, note-taking via voice helps readers to gain a better conceptual understanding of the text. This effect was explained by the difference in the content of the notes generated with the two modalities. Voice notes tended to be more comprehensive and were more elaborate than typed notes. As voice encourages readers to address a potential audience by making more person-deictic references (e.g., ``me'', ``you'') \cite{jacob2020}; it leads readers to elaborate more on the content, which, in turn, increases their conceptual understanding of the text. Second, annotating digital documents with voice notes can be faster and more convenient than textual notes. Prior research has shown that, when compared to typing using the keyboard, the act of speaking to record notes is faster and allows momentary thoughts to be recorded before they are forgotten \cite{yu2019,stifelman1993}.

Commercially available systems such as \textit{Sononcent}\footnote{\url{https://sonocent.com/}}, \textit{Luminant}\footnote{\url{https://luminantsoftware.com/apps/audionote-notepad-and-voice-recorder/}}, and \textit{Notability}\footnote{\url{https://www.gingerlabs.com/}} enables the reader to make annotations on digital documents using voice as the main input modality. For instance, \textit{Notability} is a popular note-taking application in which the reader explicitly touches the mobile display to highlight the text segments while recording a voice note about it to attach the note to it. In addition to these commercially available systems, extensions for cloud-based document viewers such as \textit{Mote}\footnote{\url{https://www.justmote.me/}} and \textit{Kaizena}\footnote{\url{https://www.kaizena.com/}} are also available, which allow readers to record voice notes on digital documents. For example, in \textit{Kaizena}, the reader first selects a text region in a single mouse movement then records a voice note to be appended to it. Further, the explicitly annotated digital document with voice notes can be easily be shared with peers. 

Even though existing note-taking applications do allow notes to be recorded using voice, these applications still rely on the reader to manually attach the recorded voice notes to the corresponding segment of text. Not only is this procedurally effortful, but it also results in the reader breaking their reading flow and diverting attention away from the written content, which adversely affects their ability to comprehend the reading material \cite{Yildiz2017}. Hence, our proposed system leverages the reader's gaze to address the challenges of anchoring voice notes to the corresponding text passage to maintain the focus of the reader on the written content.

\subsection{Gaze and Voice-based Interaction}
Although speech and eye gaze by themselves are recognised as vague input streams \cite{hatfield1997}, research shows that when combined, they could overcome each other’s weaknesses to provide enhanced multimodal interaction \cite{Kamp2011, mantravadi2009}. With recent advances in speech recognition technology, researchers have proposed a variety of systems that leverage these two input modalities for interaction~\cite{Hakkani2014, Wilcox2008, Kamp2011}. For example, Hakkani et al.~proposed a combination of gaze and voice input to enable hands-free browsing for interaction with web applications \cite{Hakkani2014}. Similarly, van der Kemp and Sundstedt proposed a drawing system controlled by voice and gaze, opening new possibilities for creative expression and giving the wider user population access to this kind of application~\cite{Kamp2011}. Further, researchers have combined gaze and voice-based interactions to enhance the immersive experience of players in games~\cite{Uludagli2018, Donovan2009}. In line with previous work, we recognise the benefit of combining these modalities and present a voice note-taking system which combines the natural behaviour of gaze while reading with the convenience and expressiveness of voice for annotating text.

\subsection{Document Annotation Using Gaze and Voice}
Whereas voice-based annotation can enable readers to take notes without turning their attention away from the written content, voice notes suffer from the problem of \textit{anchoring}, i.e.~identifying which passage in the text that the note refers to, and \textit{retrieval}, i.e.~accessing the notes at a later stage. Both are straightforward in manual note-taking but offer substantial challenges for voice notes.

The problem of anchoring relates to the challenge of correctly matching a voice note to the passages of the reading material to which it corresponds. Approaches to the task can be broadly classified as \textit{explicit} or \textit{implicit}. In explicit approaches, an item is only tagged if a user decides to associate the corresponding tags with a passage deliberately \cite{Soleymani2012}. In prior studies, researchers have explored the explicit annotation of documents, where readers interact with the application to annotate areas of documents for personal interest \cite{goularte2004interactive, Motti2009, Olsen2004}. For instance, \textit{ScreenCrayons} is a text annotating system which allows readers to perform explicit annotation on digital documents \cite{Olsen2004}. By using the system, readers can screen-capture the document and then manually highlight the relevant part of the image regarding which a written note is to be made. As a result, the reader anchors the appropriate written notes by highlighting text content.

Explicit techniques give readers full control over the annotation process, but they also cause extra cognitive effort by splitting readers' attention between annotating and reading. Implicit tagging methods attempt to overcome this limitation by tagging the data based on natural reading behaviours. Previous work using this approach has leveraged eye-tracking and other physiological data to analyse readers' visual attention in order to infer what they would like to annotate \cite{Soleymani2012}. For instance, Cheng et al.~proposed a method through which implicit annotations such as highlighting text are created when the reader looks at portions of text for an extended time \cite{Cheng2015}. In their system, a long dwell time was chosen to minimise the `Midas touch' effect when employing gaze as an input modality \cite{Jacob1990}. Similarly, implicit annotation of documents has been explored by employing the user's reading behaviour to distinguish the areas of the document that had been thoroughly read or just skimmed \cite{Vinciarelli2009, Buscher2008, Okoso2014}. For instance, Okoso et al.~employed eye-tracking to propose a feedback mechanism for second language readers. This was done by implicitly annotating the document with reading features such as the reading speed, amount of re-reading and number of fixations. \cite{Okoso2014}. 

Even though implicit gaze-based annotations open interesting opportunities for searching relevant text in the document \cite{Okoso2014}, the scope of the work done thus far has been limited to annotating passages according to the users' reading behaviour. The annotation of documents based on readers' own notes still relies heavily on the reader manually highlighting the text and linking their typed notes to it. 

To implicitly annotate the text passages with voice notes, we leverage readers' gaze. However, while recording a note, a reader may be looking at passages of a document that are unrelated to the note's content. Therefore we propose to leverage machine learning techniques to train a robust model which can learn the reading and note-taking behaviour adopted by the reader for accurately predicting the text passages a voice note is referencing.

\subsection{Machine Learning Models Utilizing Gaze Tracking}
To build a machine learning model that predicts the text passage to which the reader's voice note is referencing, we monitor the reader's reading behaviour by tracking their gaze using eye trackers. Eye trackers are now more affordable than ever and have been built into a variety of devices, from laptops to VR headsets. Previous works have demonstrated the potential of gaze data for training machine learning models for a variety of application scenarios~\cite{Biedert2012, Rello2015, Dengel2012}. For instance, a reader's English skills can be classified based on their observed reading behaviour in low, medium and high difficulty sections \cite{Yoshimura2015, augereau2016}. Similarly, Filetti et al.~proposed a PDF reader with inbuilt eye-tracking capabilities that predicts the reading comprehension by engineering features that characterise the reading behaviour of the user \cite{Filetti2019}. In line with the previous work, we engineer gaze features informed by the literature \cite{Abdelrahman2019, Srivastava2018, Cheng2015} to train a model on the reader's note-taking behaviour for anchoring voice notes to text passages.

In summary, our work aims to solve the challenge of the anchoring voice notes to text passages. For this, we draw from prior work to present a system which supports implicit annotation of documents by leveraging the reader's gaze to anchor their voice notes to text passages automatically. This supporting feature can facilitate the easy referencing of voice notes on digital documents while enabling readers to focus on the content.

\section{Study 1: Current Note-Taking Practices}
\label{Study1}

We conducted a contextual inquiry \cite{beyer1999contextual} to ground the design of GAVIN in existing note-taking practices while probing the attitudes of potential users towards the idea of a gaze-assisted note-taking application. We selected PhD students as a target user group because they frequently take notes on documents (research papers) as part of their everyday work practice \cite{Soleymani2012}. These notes are taken to summarise the findings of a paper, to make sense of technical vocabulary, or simply to capture thoughts and note down tasks to be performed at a later time \cite{Pickering2013}. Therefore, this study investigated the current note-taking practices adopted by PhD students and their opinions regarding a voice note-taking application that implicitly anchors the voice notes to text passage.

\subsection{Participants and Procedure}
We recruited 12 postgraduate research students through university mailing lists. Six participants identified as men and six identified as women. Participants were aged between 27--35 (M = 30.33 years, SD = 3.03), and were PhD students from the same department at the same university. The sample varied in terms of their research experience: five participants were in their third year of PhD study, four were in their second year, and three were in their first year.

We conducted our contextual inquiry sessions at the workspace of each participant. All observations were performed one-on-one between participants and the first author. Upon obtaining informed consent, we asked the participant to search online for a paper related to their research topic. After they had selected a research article, we asked them to read it in the manner that they ordinarily would when they were conducting a literature review. Participants were encouraged to use the editing tools provided by whichever PDF viewer they used (e.g.~Acrobat Reader) to take notes as they usually would when reading the paper. During the study, the researcher sat a few feet behind the participant and observed how participants took notes on the digital document. Where appropriate, we asked questions to probe the motivations behind their actions. For example, if a participant made an annotation on the digital research paper, we asked them to explain their reasons for making the annotation. 

Once the participant had finished reading the research article, we conducted a short audio-recorded interview in which we asked the participant to explain how they typically used their notes when drafting a literature review. We then described our proposal for a gaze-assisted voice note-taking system and inquired about the issues that the participants might have regarding the acceptance of the system. At the end of the session, we thanked participants for their time and contribution.

\subsection{Analysis}
Our data consisted of observational notes made by the researcher alongside participants' interview transcripts. The first author completed all transcription.

We analysed the data using a general inductive approach \cite{David2006}. The first author began by reading the observational notes and interview transcripts multiple times to build familiarity with the text. The same author then engaged in inductive open coding, assigning labels to the text to develop categories based on the study aims \cite{David2006}. We subsequently assigned participants' statements and observed note-taking practices to the developed categories. We then conducted a coding consistency check \cite{David2006} in which the third author deductively applied the developed categories to the raw data. This author coded the data independently, assigning the defined categories to the text. We observed a 90\% overlap between the two coders and resolved all disagreements through discussion.

\subsection{Findings}
We split the findings of the contextual enquiry into two high-level categories. The first category encapsulates the current note-taking practices adopted by participants while reading digital documents. Here, we found that participants would often associate notes with text to record essential findings while reading a digital document. These notes could be both short: \textit{``I highlight and associate small notes of one or two words such as: interesting, search more, good reference and gaps''} (P9) and comprehensive  \textit{``I usually make notes after important paragraphs briefing the key points of the paragraphs. I also write a summary note at the end of reading the paper containing the key points identified''} (P5). Further, these recorded notes support participants' later stages of work, such as when drafting a literature review. Hence, it is crucial that their notes should be anchored to correct text passages so that while revisiting notes, they could easily understand the context in which they recorded them :\textit{``I usually don't have time to reread the paper, so I would just read the notes, summary of the paper and read the highlighted section of the paper''} (P3).

The second category reports the feedback of the participants regarding the envisioned system. Overall, participants were intrigued by the idea of GAVIN. They recognised the benefits of taking notes via voice because it could save time as speaking is often faster than typing on a keyboard: \textit{``I would prefer this system, as writing with a keyboard is much more time consuming then speaking''} (P2). In addition, the notion of speaking notes aloud was seen as something that could increase focus on the task: \textit{``the system would be adaptable because speaking of what I am thinking makes me figure out my understanding. I would be less sleepy and more involved in reading the paper''} (P8). However, participants also cited a need for high accuracy in the system's ability to anchor voice notes to relevant passages of text: \textit{``I don't trust the application to make a correct annotation. I can do this process faster manually, so for me to use, it should correctly annotate my notes''} (P3). 

\textit{Implication:} The above mentioned key findings regarding the perceived benefits of voice notes and a crucial need to correctly anchor the recorded voice notes to reference text passages motivated us to propose a note-taking system, in which notes are recorded by using voice as the input modality and are implicitly anchored to correct reference text passages. We ensure that the proposed system is accurately anchoring voice notes by proposing a robust machine learning model which is trained on the gaze patterns of reader's note taking behavior to predict the passage to which a voice note refers. By presenting such a system, we facilitate the readers to easily retrieve their voice notes and understand the context in which notes were recorded.

\section{STUDY 2: Mapping Voice Notes to Text}

Motivated by the findings of the contextual inquiry, we began the development of a system that would accurately and implicitly map voice comments to text passages. To make this inference, we leverage the user's gaze behaviour as they read the text.

\begin{figure}[tbp]
  \centering
  \includegraphics[width=0.9\columnwidth]{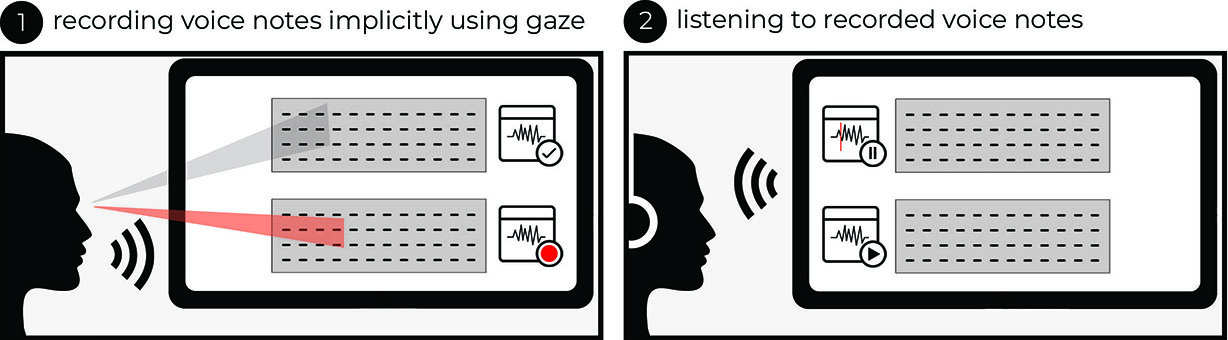}
  \caption{Proposed gaze based note-taking system for implicit anchoring and easy retrieval of voice notes. }
  \label{fig:SystemDesign}
\end{figure}

The gaze-assisted voice note-taking system we envision would allow the reader to record voice notes on digital documents. The system would then anchor the voice note to the corresponding text passage. For the retrieval of voice notes, the reader could open the article and click on the anchored voice note to listen to the recorded voice note, as shown in Figure~\ref{fig:SystemDesign}.

The first step of building this system is to understand which behavioural features allow us to anchor voice notes to text passages implicitly. We address this task by training a robust machine learning model that predicts the text passages for anchoring voice notes based on the reader's gaze activity. As note-taking is a personalized activity, readers take various kind of voice notes while reading digital documents. Therefore, to build such a machine learning model, it is essential to observe the gaze patterns which could be associated with different kinds of notes. To collect this dataset, we conducted a laboratory study in which participants were asked to read a research article. We were particularly interested in investigating the different gaze behaviours observed during voice note-taking and the usage of gaze to anchor voice notes to text passages accurately.

\subsection{Participants}
We recruited 32 participants aged between 27--42 (M = 31, SD = 3.6) for the study using recruitment posters on noticeboards around the campus. Eighteen identified as men and fourteen identified as women. All participants were PhD students from the same department at the same university. The participants were diverse in terms of the PhD year enrolment, with eight of them in their first year, eight in their second, seven in their third, and nine in their fourth year of PhD candidature. These students were also diverse in the kind of research they conducted: 13 reported working primarily with qualitative methods and 19 with quantitative methods. 

\subsection{Study Materials and Experimental Setup}
\begin{figure}
\centering
  \includegraphics[width=0.7\columnwidth]{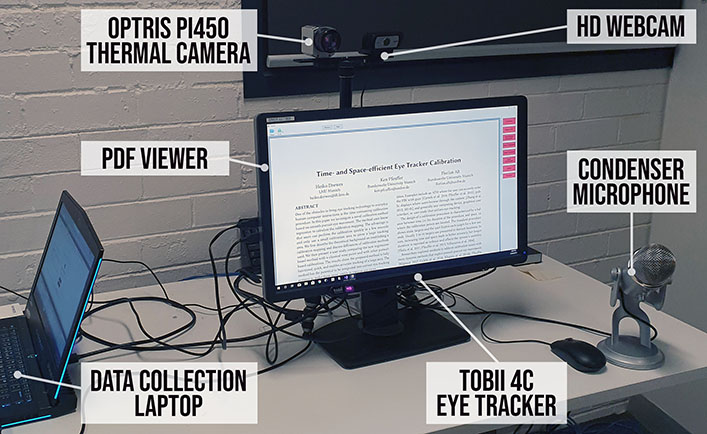}
  \caption{Experimental setup of Study 2, consisting of a thermal camera, eye tracker, web camera and microphone facing the participant. }~\label{fig:SetUp}
\end{figure}

Figure \ref{fig:SetUp} shows the experimental setup of our lab study. During the study, we recorded participants' eye movement with a Tobii 4C eye tracker (90Hz)\footnote{\url{https://gaming.tobii.com/product/tobii-eye-tracker-4c/}}, their facial temperature with an Optris PI450 thermal camera (80Hz)\footnote{\url{https://www.optris.com/thermal-imager-pi400i-pi450i}}, and their facial expression with a HD webcam. We recorded the voice of participants using a Blue Yeti condenser microphone placed on the desk, in front of the participant. The conducted study is part of a broader research project regarding understanding the cognitive effort involved in a gaze-assisted voice note-taking system. Hence, the analysis of data retrieved through the facial imaging sensors is out of the scope of this paper. 

We built a document viewer using the PDFium library\footnote{\url{https://pdfium.googlesource.com/pdfium/}} to capture changes in visual stimuli of the document seen by the participants by recording the scroll value wherever the participant scrolls the page up or down. All participants read their chosen paper using this document viewer. To record the sensor data while participants carried out the task, we built a data collection application that saved the participant's eye-tracker data, webcam images and thermal camera images.  We recorded the current scroll value and page number of the research paper that the participant was reading whenever they interacted with the scroll bar through the mouse or changed the page of the paper. We used a Blue Yeti condenser microphone to record the voice of the participant throughout the study. The application also recorded the timestamp when the participant started the experiment to synchronise the voice recording of the experiment with the sensor data. While the participant was performing the task, an additional gaze trace, not visible to participants, was captured by recording the desktop screen of the participant overlaid by Tobii Ghost \footnote{\url{https://gaming.tobii.com/software/ghost/}} by using OBS Studio \footnote{\url{https://obsproject.com/}}.

\subsection{Procedure}\label{procedure}
Upon recruitment and prior to the session, we asked each participant to send a short research paper (3--5 pages) via email that they had not previously read and which would be of interest to them. We used this paper as the participant's reading material during their study session. We chose this approach over giving them a pre-selected paper to increase the realism of the task and the diversity of formatting layouts---we observed various layouts, including single- and double-column documents, with various figure types. We asked participants to complete a demographic questionnaire that included the type of research they conducted, their stage of candidature and their native language. 

We conducted the study at our university usability lab, with each session lasting approximately 60 minutes. Upon arrival, we obtained informed consent and gave participants a plain language statement to explain the research. The main task of our study consisted of reading the paper and making voice notes about its content. We opened their chosen paper in our custom-built PDF viewer and asked participants to read the paper as if it was part of a literature review. Given that vocalising thoughts was not an existing practice amongst our participants, we began by showing examples of the kinds of voice notes that they could make while reading. We then calibrated the eye tracker with the manufacturer's default procedure. The researcher conducting the study left the room so that the participant could take voice notes without being affected by their presence. We asked participants only to speak when they wanted to make a note (though they were free to stop the study at any point) and not to make any other type of notes with other modalities, such as the mouse, keyboard, pen, and paper.

The researcher re-entered the room once the participant had finished reading the paper. Then, the researcher showed the participant their screen recording overlaid with the gaze traces to obtain the ground truth data about the relationship between voice notes and the text passages. We asked the participant to highlight the text passages in the research paper to which each voice note corresponded. To assist with recall, participants could observe their gaze traces and listen to their voice notes. The highlighted text by the participant served as ground truth for our machine learning model, described in the next section. We rewarded each participant with a gift voucher of \$20 for their involvement in the study.

\subsection{Analysis}

We analysed the data obtained from the study in two stages. We first categorised the different kinds of voice notes participants made to observe the gaze patterns associated with them. Second, based on the findings of this data analysis, we proposed a six-step method for accurately anchoring voice notes to text passages (Section \ref{Study 2}).

\subsection{Findings: Note-Taking Practices and Gaze Patterns} \label{exploratoryData}We observed that our participants' voice note-taking practices could be divided into three categories. 
\begin{enumerate}
   \item Short Voice Notes: Participants sometimes recorded short voice notes as soon as they read something important. While doing this, participants typically looked at the lines of text that had elicited the voice note. In this type of note, participants usually re-stated what they had just read.
   
     \item Reflective Voice Notes: Participants reflected on the research paper by linking content that they had just read to their prior knowledge, or to material that they had read in the past. When making these reflective notes, some participants looked at the content that had elicited the voice note, whereas others did not, instead fixating on any particular passage, which led to arbitrary gaze patterns.
 
   \item Summary Voice Notes: Some participants also recorded comprehensive voice notes, which were similar to the extensive notes taken by participants in Study 1. We observed that participants did this by reading a whole section of the paper, or in some cases, two to three paragraphs consecutively. They would then reflect on their understanding of the content by recording long comprehensive notes, summarizing the content they had just read. We observed that participants did not always record these notes while fixating at relevant passages of text.  Instead, while recording comprehensive notes, they often fixated over some text passage that was not relevant to the recorded voice note.
\end{enumerate}
 
A final observation was that the voice notes participants recorded while reading the research paper often had semantic and syntactic similarity with the text passage regarding which a note was made. This was because, when they were recording the notes, they verbalised their thoughts by using words that were in the reference text passage that they had just read.

Based on the observed note-taking behaviours and gaze patterns, we concluded that the anchoring of a participant's voice note does not always correspond to the location in which the participant was looking while speaking. Depending on the kind of note being made, it is possible that the participant might be looking at an arbitrary text passage while reflecting on their understanding via voice. This behaviour introduces considerable difficulty into the task of automatically anchoring voice notes to relevant sections of text. Therefore, it is not sufficient to simply map utterances to the text passage that the participant was looking at the time of a voice note recording. Instead, it is necessary to analyse the broader gaze patterns made by the participant before recording a note.

\section{Implicit Annotation of Text with Voice Notes}\label{Study 2}

\begin{figure}[t]
\centering
  \includegraphics[width=0.7\columnwidth]{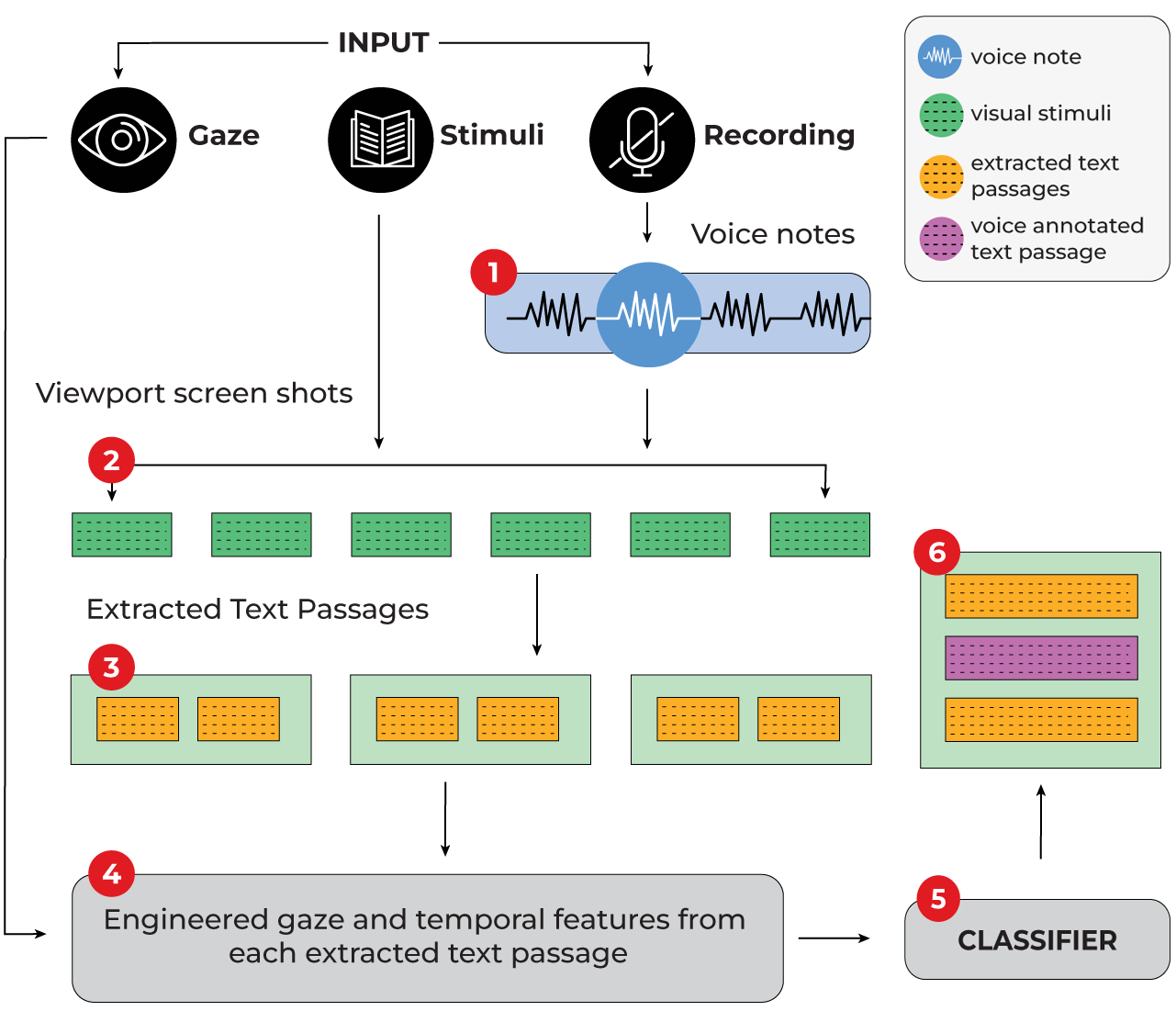}
  \caption{Proposed six-step process for anchoring voice notes to text passages. }~\label{fig:process}
\end{figure}

The findings of the exploratory data analysis informed the creation of a six-step pipeline for anchoring voice notes to text passages shown in Figure \ref{fig:process}. The input to this data processing pipeline is the raw voice recording, the recorded gaze trace, and the changes in the visual stimuli, i.e.~the scrollbar value. The output is the mapping between voice notes and text passages.

\begin{figure}[t]
\centering
  \includegraphics[width=1\columnwidth]{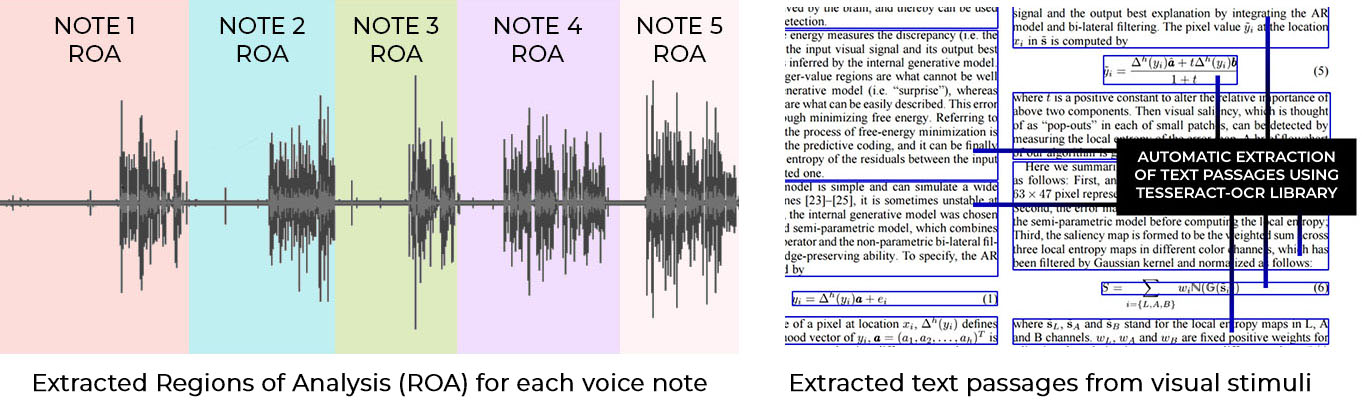}
  \caption{Left: We define a region of analysis as the time from the end of the previous note until the end of the current note. Right: We extract text passages automatically from the PDF document using computer vision.}~\label{fig:ROA}
\end{figure}

\begin{enumerate}
\item \textit{Extracting Voice Notes:} The first step in our pipeline was to process the raw audio data to extract the voice notes. We first filtered the data to remove noise with a noise reduction level of 12dB and a frequency smoothing of 3 bands. We then thresholded the signal at 26dB to remove silent segments and discarded segments under 3s. This resulted in a set of 691 separate timestamped voice notes (avg. 22 per participant). \\ \\ 
The gaze patterns we observed while participants recorded voice notes indicated that the recorded notes were not only related to the text passages where participants were looking while they spoke but were also related to the text passages participants had read before recording a note. Hence, for each voice note, we defined a Region-of-Analysis, or \textit{ROA}. This region starts from the end time of a particular voice note to the end time of the successive voice note under consideration. An example of the ROA for each voice note is as shown in Figure \ref{fig:ROA}. \\

\item \textit{Extracting Visual Stimuli:} As we do not make any assumptions about the page layout, we must use computer vision techniques to parse what the user is looking at. While participants took voice notes, we recorded information about what part of the document was being displayed in the viewport. This allowed us to fetch the image data frames within the \textit{ROA} in order to map the gaze data to pixels in the document. Our PDF viewer allowed us to replay the page navigation and scrolling together with the gaze data. \\

\item \textit{Processing Visual Stimuli:} From these extracted visual stimuli, we then used the PyTessBaseAPI \footnote{\url{https://github.com/sirfz/tesserocr}} python library to extract text passages. PyTessBaseAPI is a python wrapper for the tesseract-ocr \footnote{\url{https://github.com/tesseract-ocr/}} library, which provides functions to recognize characters from images as well as to segment images into text components. We used this library to extract paragraphs from the extracted images by giving the paragraph as an input component. An example of the extracted paragraph components from the visual stimuli is shown in Figure \ref{fig:ROA}. The extracted text passages later served as the possible passages for annotating voice notes. \\ 

\item \textit{Computing Features:} For each voice note, we pre-processed the gaze traces observed in the \textit{ROA} by first removing outliers (gaze points that lie outside the display area of the screen) and then by computing fixations. For this, we used the Dispersion-Threshold Identification algorithm \cite{Salvucci2000}, as it clusters gaze points into fixations by using only two parameters, dispersion, and duration threshold, which we set to 25 and 100, respectively. Naturally, not all fixations were within the text passages, due to calibration offsets and tracking errors. We, therefore, assigned each fixation outside the text passages to the nearest extracted passage by using hierarchical clustering \cite{Murtagh2011}. \\

After we allocated fixations to text passages, we engineered 15 spatially based gaze features and temporal features (see Table~\ref{tab:table1}). We computed the low-level gaze features for all gaze points within each text passage, which included the number of fixations in a passage normalized by the area of the passage, the maximum, minimum, and average fixation duration occurred in each text passage. The gaze features also included those engineered from analyzing the saccades occurring in each text passage. From the extracted saccades, we computed the maximum, minimum and average saccade length, duration and velocity. Beside these gaze features, we also extracted two temporal features for each text passage. These include the total time spent on reading the text passage normalized by the area of the passage and the temporal order in which each passage was read. The temporal order gave more priority to those passages that were read near the time the participant started recording a voice note. This order was calculated by taking an average of the time of the first five fixations observed in a text passage. The average time for each text passage was then subtracted from the time when the participant started recording a voice note. The obtained time for all text passages was then normalized to a value between 0-1. \\

\begin{table*}
  \caption{Engineered temporal and gaze features extracted for each text passage.}
  \label{tab:table1}
  \begin{tabular}{ll}
    \toprule
    \textbf{Temporal Features} & \textbf{Gaze Features} \\
    \midrule
    Normalized time duration for passage & Normalized fixation count \\
    Temporal order of reading passage & max, min, avg fixation duration passage \\
    & max, min, avg saccade length \\
    & max, min, avg saccade duration \\
    & max, min, avg saccade velocity \\
    \bottomrule
  \end{tabular}
\end{table*}

\item \textit{Classification Approach:} The aim of the classifier was to map the engineered features of gaze and temporal data computed from the extracted text passages to classify whether a voice note was related to a specific text passage or not. Hence for each extracted text passage either the classifier predicted \textit{Not annotated} (voice note was not made regarding this text passage) or \textit{ Annotated} (voice note was made regarding this text passage). To solve this binary classification problem for each text passage, we trained a Random Forest classifier with 1000 trees on the whole dataset of 32 participants. For each particular voice note, the number of text passages highlighted by the participants during ground truth collection  (text passages with labels \textit{Annotated}) was much smaller than the ones that were not annotated. As such, our data set was highly imbalanced. For evaluating the classifier on an imbalanced dataset, in line with previous work, we have used the metrics $F_1$-Score and Area Under the ROC Curve (AUC) (in both cases, higher is better) to assess the performance of each classifier \cite{Qiong2009}. \\

As note-taking behaviour is an individual activity that might differ from person to person, we evaluated our classification approach in two ways. We first performed a person-dependent classification by evaluating the classifier on the data from the same participant on which it was trained, but from a different note. We conducted separate evaluations for each participant, building and evaluating the classifiers using leave-one-note-out cross-validation. For example, if a participant made 15 voice notes, we trained the classifier 15 times, each time training on the 14 voice notes and evaluating it on the remaining last voice note. To build a more generic classifier, and to avoid overfitting the classifier to the voice note-taking behaviour of a particular participant, we also performed a person-independent evaluation. We evaluated the classifier by training it 32 times using leave-one-participant-out cross-validation. Each time, we trained it on the data of the 31 participants and then evaluated it on the voice notes of the last participant. The reported results, in the next sections, are averaged by the total number of participants. \\

\item \textit{Prediction of Text Passages For Annotation of Voice Notes:} The results obtained from the classifier were then used to recommend the text passages for voice annotation. For a particular voice note, the text passages which were classified as \textit{Annotated} are the proposed text passages for anchoring voice note. 

\end{enumerate}

\subsection{Baseline Approaches for Implicit Tagging of Voice Notes}
To demonstrate the performance of our approach to implicitly anchor voice notes, we implemented two baseline approaches and compared them to our proposed method.

\subsubsection{Position Data-based Voice Annotation: }
A naive approach for implicitly anchoring voice note could involve mapping the recorded voice note to the text passage, which is displayed on the top of the PDF page read by the participant. To implement this approach, we used the traces of the position data associated with the vertical scrollbar of the PDF viewer on which the digital document was bring read by the participant. 

To implicitly anchor voice notes based on the position data, for each particular voice note, we first retrieved the vertical scrollbar value and page number of the PDF document being read by the participant at the time the voice note was recorded. We then used the retrieved scroll value and the page number to extract the image data frame, which was displayed on the viewport when the participant started to record the voice note. The text passage displayed on the top of the extracted image frame is classified as  \textit{Annotated}, and the recorded voice note is anchored to it. Further, we classified all other text passages on the extracted image frame as  \textit{Not Annotated} for the particular voice note. 

\subsubsection{Gaze Fixation-based Voice Annotation: }
Another approach for implicitly annotating digital document with voice notes involves mapping the voice note to the text passage participant has looked the most while speaking. To implement this approach, for each voice note, we analysed the gaze trace observed during a window ranging from the start time of the participant’s utterance to its end time. We first clustered the gaze points observed in this window into fixations using the Dispersion-Threshold Identification algorithm \cite{Salvucci2000} by setting the value of dispersion and duration threshold parameter to 20 and 100, respectively. The obtained fixations were then assigned to the nearest extracted text passage. We then counted the number of fixations within each extracted text passages residing in the time span of recorded voice note. The text passage which had the highest count of gaze fixations was classified as \textit{Annotated} whereas all other extracted text passages were classified as \textit{Not Annotated}.

We evaluated the two baseline methods on the collected dataset and reported the performance in terms of $F_1$-Score and AUC metrics. The reported results in the next section are averaged by the total number of participants.

\subsection{Results}
We first report and compare the performance of the person-dependent and person-independent classifiers for voice annotation. We then compare our proposed approach with the two implemented baseline approaches for implicitly annotating digital documents with voice notes. Further, we explore the reasons why certain gaze and temporal features have more impact on the performance of the participant-independent classifier. Finally, we discuss the cases where our classifier failed to anchor the voice notes to the correct passage.

\subsubsection{Performance:} We computed the precision, recall, $F_1$-Score and AUC to evaluate the performance of the classifier. Our results show that it is feasible to anchor a voice note to the appropriate section of text by using spatially based gaze features and temporal features (see Table ~\ref{tab:table2}). Our classifiers achieved an average $F_1$-Score of \textit{0.84}, which is sufficient to implicitly anchor the reader's voice notes to the correct text passage. 

\begin{table*}
  \caption{Participant-dependent and participant-independent classifier performance.}
  \label{tab:table2}
  \begin{tabular}{lcc}
    \toprule
    \textbf{Evaluation Measures} & \textbf{Person Dependent} & \textbf{Person Independent} \\
    \midrule
     Precision & 0.87 $ \pm $ 0.07 & 0.79 $ \pm $ 0.14 \\
     Recall & 0.90 $ \pm $ 0.10 & 0.83 $ \pm $ 0.15 \\
     $F_1$-Score & 0.88 $ \pm $ 0.08 & 0.79 $ \pm $ 0.11 \\  
     AUC & 0.94 $ \pm $ 0.04 & 0.89 $ \pm $ 0.07 \\
    \bottomrule
  \end{tabular}
\end{table*}

\begin{table*}
\caption{Performance of the person-independent classifier for implicitly anchoring different kind of voice notes to text passages.}
  \label{tab:table4}
  \begin{tabular}{lcc}
    \toprule
    \textbf{Type of Voice Notes} & \textbf{AUC} & \textbf{$F_1$-Score} \\
    \midrule
     Short Notes & 0.93 & 0.82  \\
     Reflective Notes & 0.85 & 0.76 \\
     Summary Notes & 0.92 & 0.81 \\  
    \bottomrule
  \end{tabular}
\end{table*}

We observe that the person-dependent classifier could more accurately classify the text passages that were labelled as \textit{Annotated} as compared to the person-independent classifier (see Table \ref{tab:table2}). This is because the person-dependent classifier is trained and evaluated on the data of the same participant; hence the trained model can overfit to the note-taking behaviour of the same participant and give average high AUC for that participant. This result implies that the approach of using a person-dependent classifier is not feasible to anchor the text passage with voice notes made by a new user. In contrast, although a person-independent classifier has a relatively lower AUC of \textit{0.89}, it still generalizes well to the new user and is precise enough to anchor voice notes with text passages for any new user. 

Moreover, to identify the note-taking patterns on which the person-independent classifier could most accurately predict the text passages for anchoring voice notes, we tested the model on the three different kinds of voice notes made by the participants that were identified in the exploratory data analysis (see Section ~\ref{exploratoryData}). We observed that out of 320 \textit{short} voice notes, 230 (72\%) were annotated correctly. From the 144 \textit{reflective} voice notes, 91 (63\%) were anchored correctly. Lastly, out of a total of 95 \textit{summary} notes, (66\%) voice notes were annotated to correct text passages. We further report the classification performance to annotate text passages for the three kinds of voice notes in Table \ref{tab:table4} using the $F_1$-Score and AUC metrics. 

The results in Table \ref{tab:table4} indicate that the person-independent classifier could generally understand the note-taking patterns exhibited by all the different kinds of voices notes recorded by the participants. However, compared to the reflective voice notes, the classifier more accurately predicted labels for text passages for short and comprehensive voice notes. These results were expected as participants tended to look towards to the text passages they were speaking about while recording short and comprehensive voice notes. Compared to this behaviour, while recording reflective voice notes, participants often related the read concepts with their prior life experience; hence exhibiting arbitrary gaze patterns, which contributed to relatively lower classification performance.

\begin{table*}
\caption{Performance of position data-based and gaze fixation-based baseline approaches for implicitly anchoring voice notes to text passages.}
  \label{tab:table3}
  \begin{tabular}{lcc}
    \toprule
    \textbf{Evaluation Measures} & \textbf{Position Data} & \textbf{Gaze Fixation} \\
    \midrule
     Precision & 0.15 $ \pm $ 0.12 & 0.69 $ \pm $ 0.27 \\
     Recall & 0.48 $ \pm $ 0.41 & 0.43 $ \pm $ 0.23 \\
     $F_1$-Score & 0.48 $ \pm $ 0.48 & 0.48 $ \pm $ 0.19 \\  
     AUC & 0.58 $ \pm $ 0.27 & 0.69 $ \pm $ 0.10 \\
    \bottomrule
  \end{tabular}
\end{table*}

\subsubsection{Comparison with Baseline Approaches:}

\begin{figure}[t]
 \centering
 \includegraphics[width=0.7\textwidth]{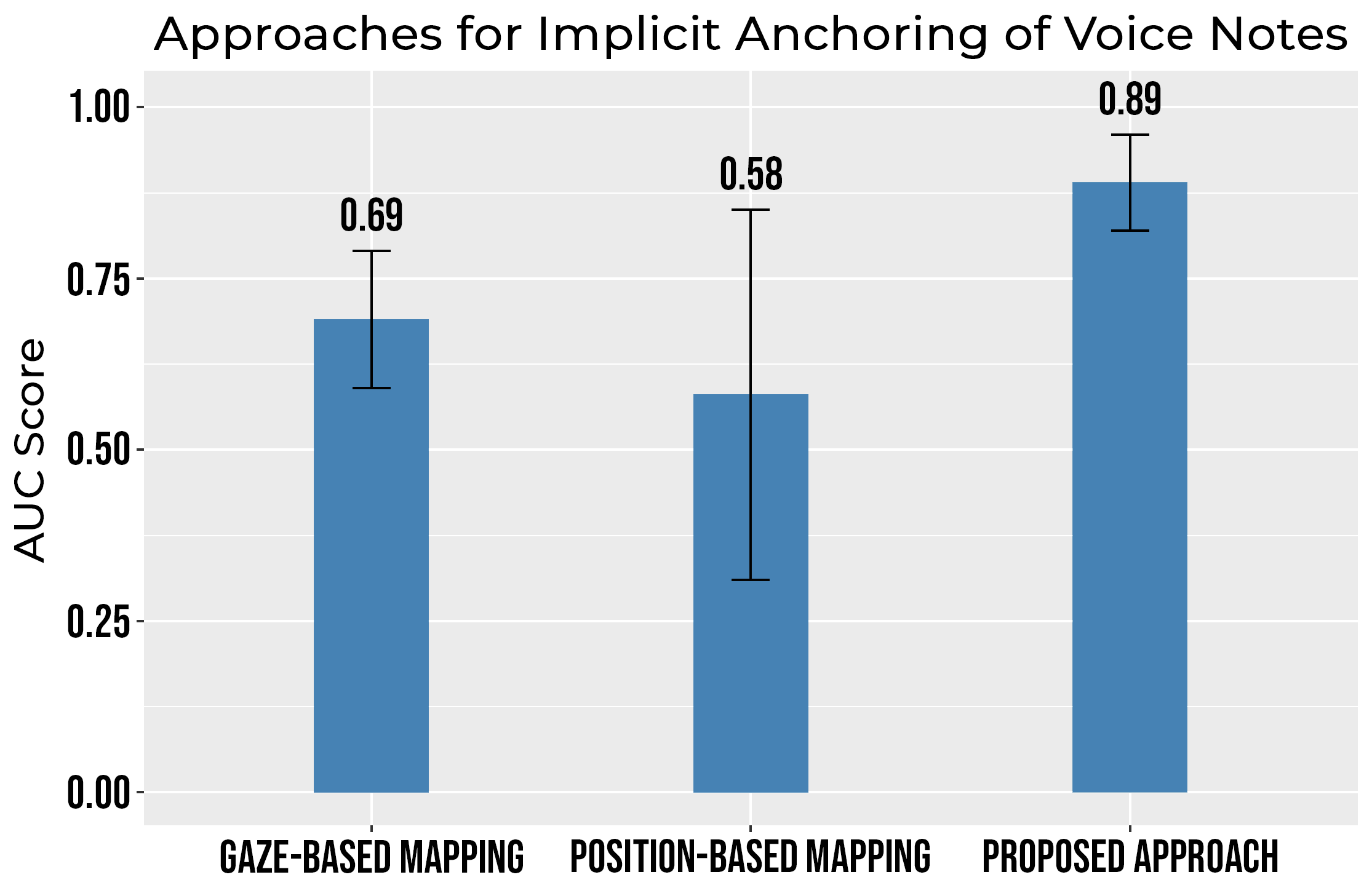}
 \caption{Comparison of our proposed approach employing gaze and temporal features with two implemented baseline approaches for anchoring voice notes to digital documents. Error bar represents standard deviation.}
  \label{fig:Comparison}
\end{figure}

The results in Table \ref{tab:table2} show the performance of the two baseline approaches leveraging the position data of the digital document and participants' gaze trace to anchor voice notes implicitly. The results indicate that our proposed approach of a person-independent classifier trained on spatially based gaze and temporal features outperformed the two baseline approaches by an average of \textit{25\%} for the AUC score (see Figure \ref{fig:Comparison}). This increase in performance could be explained by the observations of our exploratory data analysis (see Section ~\ref{exploratoryData}) that mapping of a voice note does not always correspond to the location at which the participant is looking while speaking, or to the changes in vertical scrollbar value made by the participant while recording a voice note. Depending on the different note-taking patterns, the participant might be looking at an arbitrary text passage or scrolling the document while recording a voice note. Therefore, our proposed approach capturing the broader reading and note taking patterns is more effective in anchoring voice notes to text passages.

\subsubsection{Feature Importance:} 
\begin{figure}[h]
\centering
  \includegraphics[width=0.7\textwidth]{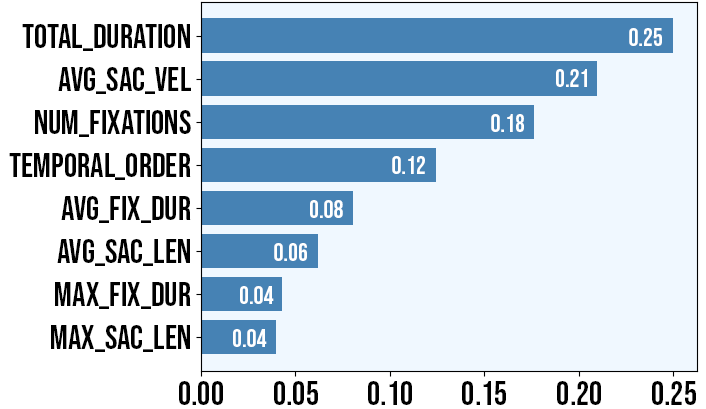}
  \caption{Top eight important features for anchoring voice notes with text passages.}~\label{fig:ImportanceFeatures}
\end{figure}

We further investigated the importance of the features used for training the participant-independent classifiers. We used the Random Forest importance score to observe the top-eight important features to classify whether a particular voice note is made regarding an extracted text passage based on gaze and temporal features. As seen in Figure \ref{fig:ImportanceFeatures}, it was observed that no single feature is entirely responsible for the classification task. However, a combination of temporal, fixation and saccade-based gaze features are significant in deciding whether a voice note was made regarding a passage. We observed that the two temporal features---time spent on reading a text passage and the temporal order in which a passage was read---received high importance. This is because the passage regarding which a note was made has some critical content for the participant to make a voice note; hence, the time spent on reading this passage was often more than other passages. 

Similarly, we also observed that participants made voice notes as soon as they read some text which seemed important to them.  Hence, the temporal order played a significant role in discriminating between text passages that were read with a similar gaze pattern. For example, if two text passages were read with comparatively similar gaze patterns, but one of them was read just before the time when the participant started recording a note, the classifier would predict the later one in time order.  Low-level gaze features such as the normalized count of fixations and average saccade velocity and length occurring in a text passage also got high importance, in line with previous work (e.g.~\cite{Dengel2012, kunze2013}). Thus, we can conclude that those text passages which participants read comparatively more carefully to make a voice note resulted in a high number of fixations, relatively less saccade distance and low velocity of horizontal saccades.

\subsubsection{Need for Textual Modelling of Comments: } \label{ModellingNeed} Although the generic person-independent classifier has an AUC score of \textit{0.89} for anchoring voice notes with text passages, the classifier in a few cases predicted \textit{Not Annotated} labelled text passages as \textit{Annotated}. This resulted in a relatively lower precision and recall score for the person-independent classifier. 

\begin{figure}[h]
\centering
  \includegraphics[width=1\textwidth]{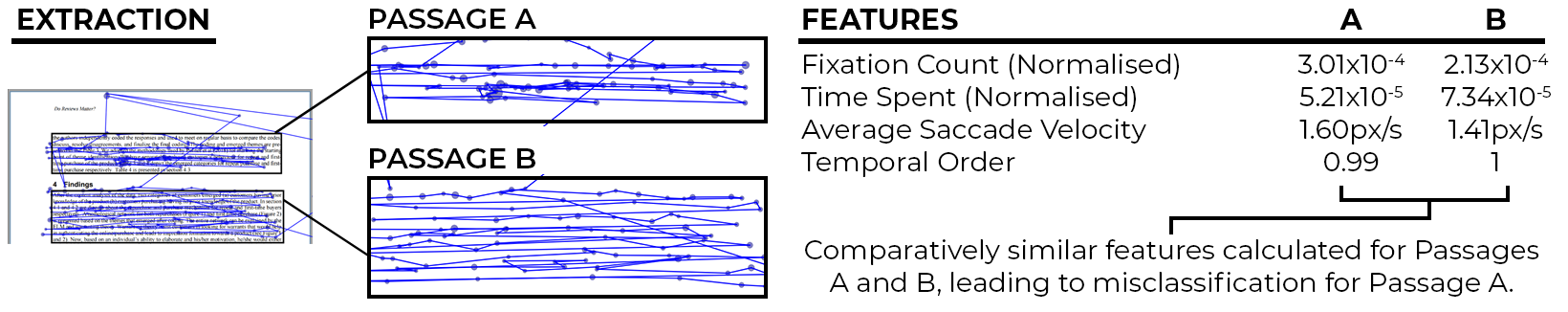}
  \caption{Observation of comparatively similar features across Passage A and B, leads misclassification to Passage A.}~\label{fig:fail}
\end{figure}

Figure \ref{fig:fail} shows an example of misclassification, where Passage B is the actual passage regarding which a voice note was made. The trained classifier, in this case, would predict both passages --- Passage A and Passage B for voice annotation. This is because similar gaze patterns are observed in both text passages, and they both are read very near the time when reader started recording the notes. In all such cases, the trained classifier would predict more \textit{Not Annotated} text passages as \textit{Annotated} for voice note annotation. One possible way of solving these tiebreaker cases would be to analyze the text of the transcription of the voice notes made by the reader. 

\section{GAVIN: Gaze-Assisted Voice-based Implicit Note-Taking System}
Based on our findings obtained from observing the users' note-taking practices (Study 1) and using proposed classifier (Study 2), which can accurately predict ($AUC=$ \textit{0.89}) the passages for annotating voice notes we implemented GAVIN.  Our prototype is a voice note taking-application which implicitly anchors the recorded voice notes with reference text passages while the user reads an article. Using GAVIN, the user can open any PDF document and record voice notes on it. The notes, based on the prediction of the annotation classifier (see Section \ref{Study 2}), are then anchored to the reference text passages inferred from their gaze behaviour.

\begin{figure}[h]
\centering
  \includegraphics[width=1\textwidth]{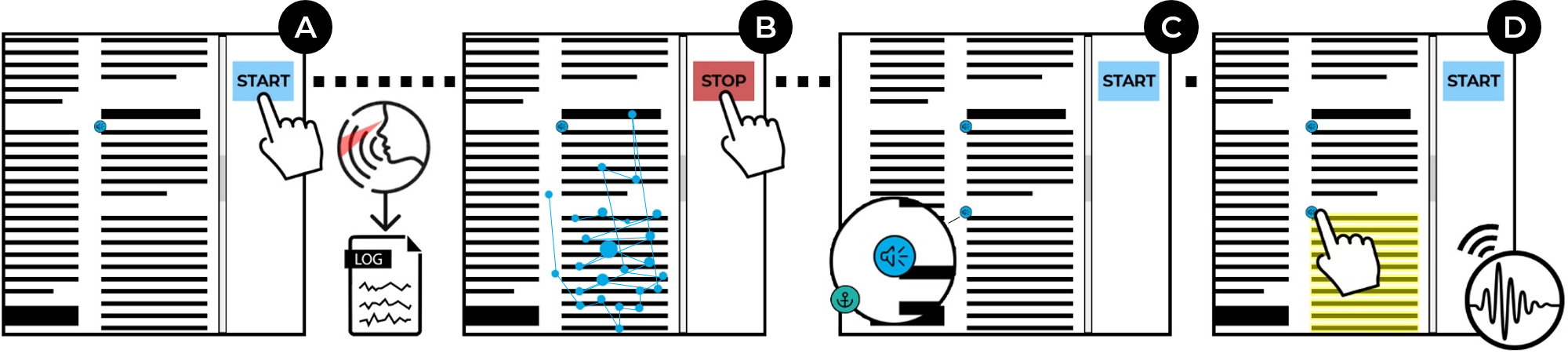}
  \caption{\textbf{A}: The Reader presses the Start button to begin recording a voice note. \textbf{B}: Reader stops the recording by pressing the Stop button. \textbf{C}: A sound icon is added to the top of the predicted text passage. \textbf{D}: Reader retrieves the voice note by tapping on the sound icon. This action plays the voice note and highlights the corresponding text. }~\label{fig:software}
\end{figure}

To develop GAVIN, we created a PDF Viewer application embedded with eye-tracking technology using the PDFium.Net library in C Sharp. The Tobii 4C eye tracker was used to monitor the gaze activity of the user. Upon loading the document in the viewer, text passages of the PDF documents were extracted using the PyTessBaseAPI \footnote{\url{https://github.com/sirfz/tesserocr}} python library. The system records the gaze coordinates while the user reads and records voice notes. When a voice note is complete, the gaze coordinates are mapped to the page coordinates to keep track of where the user was looking on the PDF page while reading. These page coordinates are then allocated to various extracted text passages, and region-based gaze and temporal features (see Table~\ref{tab:table1}) are calculated. The engineered features are further fed into the pre-trained person-independent classifier to classify each text passage into \textit{Annotated}  or \textit{Not Annotated} class. Based on the prediction made by the classifier, a sound icon is anchored to the top of the predicted reference text passages. The two actions that have been defined for the note-taking system are shown in Figure ~\ref{fig:software} and described below:

\begin{enumerate}
   
\item \textit{Anchoring Voice Notes:}
To record a voice note while reading an article, the user has to press the recording button located at the left side of the viewer (see Figure ~\ref{fig:software}) and speak out loud their notes. A prediction is made by the classifier regarding the reference text passage based on the user’s gaze activity. The voice note is saved in the system's memory, and a sound icon is displayed beside the reference text passage.

\item \textit{Retrieval of Voice Notes:}
The user can retrieve the recorded voice note and visualise the anchored text passage by tapping on the sound icon displayed on top of the reference text passage (see Figure ~\ref{fig:software}). This action plays the recorded voice note as well as highlights the reference text passages. 
\end{enumerate}

\section{STUDY 3: USER EVALUATION OF GAVIN}

We conducted a feasibility study of the voice note-taking system to gather user’s feedback. Our study was structured to probe the potential opportunities and limitations of the possibilities afforded by our technical prototype. Specifically, the two research objectives we wanted to investigate were:

\begin{enumerate}

\item What is the perceived usefulness of the voice note-taking application that implicitly anchors the voice notes to the corresponding content?

\item What obstacles the users might face in using the system and what are the ways through which we can overcome these obstacles?

\end{enumerate}

\subsection{Participants }

We recruited 12 participants aged between 24–33 (M = 29, SD = 2.9) for the study through the university mailing lists. Five identified as women and seven identified as men. All participants were PhD students from the same university.

\subsection{Study Materials and Experimental Setup}

The experimental setup for this study is in accordance to Figure ~\ref{fig:gavin}, where a participant freely holds a tablet enhanced with eye tracking capabilities while using GAVIN. We deployed GAVIN on a Microsoft Surface Pro 4 tablet running Windows 10. During the study, we recorded the participants’ eye movement with a Tobii 4C eye tracker (90Hz) and voice with the inbuilt microphone of the tablet. We selected 12 PDF articles, one for each participant to read related to general topics ranging from health to education. The selected documents had various layouts, including single- and double-column and included figures and tables. 

\subsection{Procedure}
We conducted the study at our university usability lab, with each session lasting approximately 40 minutes. Upon arrival, the researcher obtained informed consent and gave participants a plain language statement to explain the research. The researcher started with first demonstrating to the participants how to use the voice note-taking system. The participants then used the system for reading the provided article and made various voice notes regarding its content. After the participants finished reading the paper, the researcher asked them to verify whether anchoring of the voice notes has been done accurately. The researcher then asked participants to rate the system on its ease of use for making voice notes and perceived usefulness on a 5-point scale. Before concluding the session, the researcher conducts a short audio-recorded interview in which the participant is asked them to comment on the potential uses of the system in an academic setting. They were also free to make suggestions on improving the system, including any features that should be added that would have enhanced their experience.

\subsection{Findings}
We analysed the interview transcripts of the participants to report the user feedback of the system by following the general inductive approach, as adopted in Study 1 (see Section ~\ref{Study1}). The system was overall well-received by our participants, and the major findings are focused on the perceived usefulness of the system and the key issues which the participants faced while interacting with the system. 

During the study, we observed that GAVIN accurately anchored participants' recorded voice notes with reference text passages. Participants made a total of 132 voice notes (avg. of 11 per participant).  We split the voice notes into three categories: (1) notes that were anchored to the correct piece of text, (2) notes that were anchored to the correct piece of text and additional incorrect text, and (3) notes that were anchored to the incorrect piece of text. Out of these, the system anchored 99 voice notes (75\%) to correct text passage, 28 voice notes (21\%) to the correct text passage and one additional incorrect passage and only 5 of the voice notes (4\%) to an incorrect passage. These promising results validate the ability of GAVIN to anchor voice notes to correct text passages.

\subsubsection{Perceived Usefulness of GAVIN}

The system was overall well received by the participants. They rated the ease of anchoring at 4.08 (SD = 0.9) and retrieval of voice notes at 4.4 (SD = 0.6). They further rated the usefulness of the system at 4.17 (SD = 0.83). We divide the feedback related to the perceived usefulness into four broad categories described below.

\begin{enumerate}
    \item \textbf{Accurate Anchoring of Voice Notes}: Participants liked the feature of voice notes being implicitly anchored to the correct text passages: \textit{``The quality of anchoring was good, I didn\textquotesingle t do much work to anchor notes, just pressed the button''} (P6). They further commented that through implicit anchoring, the need for them \textit{``to switch to another application to record a voice''} (P9) or \textit{``to drag and drop the sound recording to the text segments''} (P7) would be diminished.
    
    \item \textbf{Retrieval of Voice notes with Text Passages}: Six participants liked that while replaying the voice notes the reference text passages was highlighted. This makes it \textit{``easy''} (P5) and \textit{``convenient''} (P8) for them to look at the highlighted text and remember the context in which the note was recorded: \textit{``This feature is useful because when I press the sounding icon, it highlights the text passages so everything just maps together, the voice note and the context''} (P5). 
    
    \item\textbf{ Saving Time and Effort}: Participants agreed that taking notes using GAVIN would be faster and would require minimum effort (P1, P4, P6, P9, P11, P12). This was mainly due to two reasons. Firstly, notes were recorded via voice which is \textit{``fast and easy''} (P1, P12) than writing. Secondly, participants considered that automatic anchoring of voice notes \textit{``saves times and effort''} (P6) which further allows them to read and annotate documents in a mobile environment: \textit{``I could use this system for reading while I am writing in a tram because not much hand movement is required''} (P4).
    
    \item \textbf{Similarity with Existing Systems}: Some participants perceived that as the interface of the system for capturing, displaying and retrieval of voice notes is similar to the existing PDF readers which allow textual annotation; thus they could adapt more easily and quickly to the proposed system (P7, P10): \textit{``The interface have somewhat similarity to the existing textual note taking system, so adaptation is also easy''} (P7). 
    
\end{enumerate}

\subsubsection{Improvements for GAVIN}
While participants recognized the benefit of using GAVIN for annotating text with voice notes, they also pointed out some issues and concerns. Moreover, they also provided suggestions on improvements and additional features to be included in GAVIN to enhance their reading experience.
\begin{enumerate}
    \item \textbf{Hands-Free Recording of Voice}: Some participants were less enthusiastic about recording voice by pressing a button (P2, P6, P8, P10). They reported it to be a bit \textit{``cumbersome''} (P2). Participants further suggested, to introduce a more hands-free mean of recording a voice such as \textit{``using gaze to press a button''} (P4) or \textit{``automatic detection of voice to capture the note''} (P8). 
    
    \item \textbf{Usage in Shared Environment}: Two participants expressed their concerns about using GAVIN in a workplace context (P1, P12). They commented that recording notes via voice could disturb fellow students: \textit{``if you are working in an environment where people are around, then using this system could bother them''} (P12).
    
    \item \textbf{Transcription of Voice Notes}: As expected, some participants were concerned regarding the limitation of voice notes of not being editable (P2, P1, P4, P7, P10, P12). To overcome this, they suggested to transcribe voice notes and present it textual form, this would make their notes \textit{``editable''} (P7) and \textit{``searchable''} (P4) for later referral. Further, one participant also perceived that text transcription would make the voice notes more understandable for other readers listening to the recording: \textit{``The problem is that I have an accent which other people cannot understand clearly, so it\textquotesingle s better if my voice notes are changed to text so another reader could understand what I intend to say''} (P7).
    
    \item \textbf{Suggestions for Additional Features}: Participants suggested some additional features to be included in GAVIN to enhance their reading and annotating experience. Some suggested to make voice notes more accessible, e.g., by organizing them (P2, P9): \textit{``If we have a list of notes at the left panel of the reader … I could visualize that how many total notes I made while reading this paper''} (P3) or by using voice to re-find a particular note: \textit{``I could give a command to the system to retrieve notes from this page''} (P2). Others commented that for GAVIN to become a \textit{``complete''} (P5) annotation tool, an additive feature which facilitates selecting specific key points in the annotated reference text passage (P6, P7, P11) should be added. They commented that this would give the reader the ability to make a general note regarding the passage as well as identify major key points from the anchored paragraph: \textit{``the system should give the ability to highlight certain words ... I don\textquotesingle t want to read the whole highlighted paragraph … by using this feature, I could just listen to the voice note and look at the specific points I have highlighted''} (P5).
    
\end{enumerate}

\section{Discussion}
In this paper, we have presented GAVIN, an interactive gaze and voice note-taking system which implicitly anchors reader’s voice notes to reference text passages. GAVIN aims to make the task of annotating digital documents via voice easier and less effortful, allowing readers to maintain their focus on the written content while making voice notes.

We have taken three crucial steps toward building the system. First, we conducted a contextual enquiry to ensure that our proposed system is grounded in realistic practices. The study showed that PhD students, as an ideal target group for voice note-taking applications, regularly take notes, but using different strategies. Importantly, they were interested in using voice notes but expressed a desire for the system to be low-effort and highly accurate. Second, inspired by the contextual inquiry, we collected a dataset to train a classifier for anchoring readers' voice notes to text passages. We proposed a method that leverages machine learning applied to gaze data. Through this method, we achieved a high AUC for the trained classifiers, meaning that the classifiers had a high ability to discriminate between text passages which corresponded to voice notes and text passages which did not, validating our approach. By using just the gaze data, we were able to anchor the majority of readers' voice notes accurately. This step is essential and necessary for addressing the underlying challenges of anchoring and retrieving voice notes. Finally, we deployed the classifier in a system prototype, which we evaluated in a user study. In the following sections, we discuss the observations and issues that will inform our future work.

\subsection{Classification Performance}
To accurately anchor voice notes to text passages, we built two classifiers, \textit{person-dependent} and \textit{person-independent}. We observed that the former classifier had a higher AUC than the latter. This result is as expected, as making notes while annotating documents is a personalised activity and how people read and take voice notes varies. However, we can use each of the classifiers in different settings, and each one offers a different direction for design. The person-dependent classifier could be used to build a personalised annotating system for a specific person, as this classifier can improve its performance by learning from a person's note-taking behaviours. On the other hand, we could employ the person-independent classifier in a note-taking system used by a general audience, e.g.~a classroom where students change from year to year. 

Moreover, during the user study of the GAVIN (Study 3), we observed that the classifier occasionally misclassified text passages when readers exhibit a skimming pattern for passages that are to be annotated. An example of this observation is when reader carefully reads (characterised by a high number of fixations, long fixation durations, relatively low saccade distance) a text passage and then barely skims the successive passage while recording a short voice note regarding this passage. In this scenario, the classifier misclassifies and predicts the former passage to be annotated. A possible explanation of this behaviour is that during the training process the classifier learns that annotations are made regarding passages which user actually read rather than those passages which they have just skimmed. This observation re-instated the need (see Section ~\ref{ModellingNeed}) to analyse the textual content of the voice note to increase the accuracy of the proposed classifier. Thus, in our future work, we plan to provide automatic speech transcription of the reader’s voice notes. This feature would not only make the voice notes more accessible for readers but also help us in further increasing the robustness of the system by engineering textual features which capture the semantic similarity between the recorded notes and gaze informed text passages.

Further, we observed that the generic person-independent classifier could correctly classify text passages, that are bounded in regions with a mean width of 560px (SD = 260) and mean height of 136px (SD = 155). The performance of the classifier could further be improved to annotate text passages which are bounded in relatively smaller regions by analyzing the speech data of the voice note in combination with the proposed gaze and temporal features. We observed during the exploratory data analysis of Study 2 (see Section ~\ref{exploratoryData}) that participants fixate over regions of their interest while they record voice notes. Consequently, the recorded voice notes often include referential terms (such as \textit{"These results are important for analysis"}) which explicitly refer to their region of interest. Thus, to improve the accuracy of the classifier for smaller text regions, in addition to engineering features which capture the semantic similarity between voice notes and reference text passages, we further plan to analyze the speech content with the gaze trace to identify the text region which the reader was explicitly referring while recording a note.

\subsection{Comparison With Existing Systems}
By proposing GAVIN, we have attempted to improve the functionality of currently used voice note-taking applications such as \textit{Kaizena} and \textit{Notability}. These applications allow readers to annotate on the digital document by using both textual and audio format. However, during the task of anchoring of notes, readers have to explicitly interact with the digital document either by touch or mouse to select the text region which is to be linked with the voice or typed note. The task of manually selecting text for annotating digital documents on mobile devices could be challenging for readers due to the difficulties introduced by the ``fat finger'' problem \cite{holz2010}, and because it breaks the reading flow and diverts the reader's attention from the primary task of reading and note-taking. 

GAVIN provides a more convenient and implicit solution for annotating digital documents with voice notes. By leveraging the reader's gaze, the system predicts the text regions which are to be linked with the recorded voice notes. This implicit mapping of voice notes will make the process of annotation less effortful for readers as it diminishes the need to select text regions manually. Further, the seamless anchoring of voice notes to text passages may also assist readers in performing a more focused reading.

\subsection{ Towards Better Understanding and Encoding of Reading Material} 

The current iteration of GAVIN (presented in Study 3) requires readers to initiate a voice note recording by pressing a button displayed on the user interface. We envision to make this process more effortless by introducing hands-free approach for recording voice notes. The rapid advancements in speech recognition technology have made voice input, not only hands-free but also efficient way of interacting with mobile interfaces. Applications exist such as \textit{Click by Voice} \footnote{\url{https://github.com/mdbridge/click-by-voice}}, \textit{HandsFreeChrome}\footnote{\url{https://www.handsfreechrome.com/}} exists in which user control the application interfaces by giving voice commands. We envision to introduce voice recognition in GAVIN, through which readers could record their voice notes by giving voice commands, similar to existing intelligence voice assistant today.

Moreover, the motivation behind this hands-free mean of recording notes comes from our observation during the data collection process (Study 2, Section ~\ref{procedure}) during which participants recorded voice notes without explicitly interacting with the PDF viewer. We observed that the act of speaking out loud to record notes compels the participants to make notes in which they were explaining concepts and ideas to themselves. This act of self-explanation through verbalisation, based on the model of Berry \cite{Berry1983}, could further help in comprehending reading material. Berry concluded through a set of experiments that people who explain ideas to themselves understand concepts more deeply in comparison to those who do not \cite{Berry1983}. Similarly, Forrin and Macleod also found that self-explanation of concepts through verbalisation helps in efficiently encoding the content into readers’ long-term memory \cite{Forrin2017}. Further, the process of self-explanation can also help the readers to realise the critical ideas explained in the document they did not wholly understand, commonly known as the ``illusion of explanatory depth'' \cite{Rozenblit2002}. This realisation could then compel readers to confront their illusion and gain a more in-depth explanation of concepts. 

By recognising the benefits of self-explanation through verbalisation, in future work, we plan to introduce voice input as a mean of recording notes, so readers could record notes just by speaking out loud their notes. This capability not only makes voice annotation easier and more natural but also helps in the better encoding of content in memory through self-explanation.

\subsection{Application Scenarios}

Current annotation tools allow users to annotate digital documents by explicitly mapping digital text with notes. GAVIN extends the functionality of these applications by supporting implicit anchoring of notes, thus making the process of annotation more convenient for users. The proposed system could be used in various settings, such as those described below.

\textbf{\textit{Feedback tool.}} GAVIN could be used by instructors for grading papers and providing feedback to students. Prior research on comparing the feedback modality (Voice vs.~Text) has shown that instructors provide richer feedback in less time while using voice than compared to text \cite{ice2007, sipple2007}. The difference arises primarily because the act of speaking is much faster than typing, which allows instructors to give more detailed feedback using voice. Moreover, voice feedback is characterised by more information and richer language \cite{dagen2008}, making it well received by students. Studies suggest that compared to text, voice feedback is more meaningful and better understood by students as it contains more explanations \cite{cavanaugh2014, keane2018} and students engage with the instructor at a more personal level when feedback is provided by voice \cite{susanne2014}. We believe that GAVIN will allow instructors to capitalise on the benefits of voice feedback in a way that is less effortful than existing tools. 

\textbf{\textit{Note-taking tool.}} GAVIN could be built into a complete note-taking application which allows students to read digital documents and take contextualised voice notes. Note-taking is a common reading strategy which students often undertake to increase text comprehension and encode material in long-term memory. Prior literature indicates that students tend to ponder and elaborate more on the reading material when recording voice notes as compared to typed notes \cite{khan2020}. Hence, we believe that using GAVIN could not only assist students in recording voice notes conveniently but also assists them in gaining a higher conceptual understanding of the digital text.

\textbf{\textit{Usability testing tool.}} When designing and running usability testing sessions, it is common for researchers to employ the `think-aloud' method where participants verbalize their thoughts while they perform tasks on a system (or user interface). Current eye-tracking usability software tools often require researchers to manually determine what is being referred on display as participants think-aloud. With GAVIN, verbalized comments can be automatically anchored onto areas of interest on the interface, which then allows researchers to identify parts and functions that need to be improved conveniently.

\subsection{Limitations}

We acknowledge a few important limitations in this work. First, the studies took place in a controlled lab setting, in which participants were not allowed to exhibit otherwise natural behaviours, for example, manually highlighting or typing their notes. Participants were only allowed to take voice notes and use the mouse to navigate through the article. In the future, we plan to evaluate the next iteration of GAVIN in a more realistic setting by removing the limitations mentioned above.

Secondly, in our study, we collected the ground truth while training our model (i.e.~the actual text passage for voice annotation) after participants had completed the task. During this process, we relied on the participants' memory by replaying them their gaze trace with their voice notes and requiring them to recall which text passages they were referring to while they were recording the voice notes. Consequently, we limit the validity of our ground truth collection approach to the participant's memory. Given that asking participants to report the ground truth as they completed the task would generate gaze artefacts, we had to balance the naturalism of the behaviours with the quality of the data.

Thirdly, our data collection only involved a specific study group of people (PhD students) who were first-time voice note-takers and were allocated a specialised task of reading a research paper to conduct a literature review. As such, we trained the classifier for anchoring voice notes with text passage on the behaviour of the particular group performing a specific task. This approach is suitable for showing the feasibility of anchoring voice notes by leveraging readers' gaze; however, it is essential to verify whether the proposed approach can be generalised for a diverse audience from different educational and linguistic backgrounds. However, we expect that the more experienced users become with the system, it would lead users to exhibit behaviours that ``help'' the classifier, such as fixating on the correct text passage, which should improve the overall classification performance.

Finally, we evaluated GAVIN in a stationary environment where participants recorded voice notes while reading from a display with desk support. This evaluation approach is suitable for showing the feasibility of leveraging readers' gaze for anchoring voice notes. Future work should evaluate and compare GAVIN with other explicit approaches (e.g., touch input) of anchoring voice notes in a mobile environment where users can hold the device in different ways.

\section{Conclusion}
In this paper, we presented a gaze-assisted voice note-taking system through which readers can anchor voice notes with text passages with ease. To build such a system, we proposed a data processing pipeline that leverages the gaze activity of the user and machine learning techniques to facilitate the anchoring of voice notes made on the digital text. We evaluated our approach by training a classifier which leverages gaze features from the reading and note-taking activity of 32 participants. The results obtained showed that readers' gaze patterns do indeed correlate with their note-taking activity. We further used our trained classifier to develop an interactive gaze assisted voice note-taking system which seamlessly anchors readers' voice notes with text passage for later referral. Lastly, we conducted a user study on the voice note-taking system, which provides important insights and guidance for future work. We further envision that the built system would enable readers to perform focused reading and polish their metacognitive skills by speaking out loud their understanding of the reading content.

\begin{acks}
Eduardo Velloso is the recipient of an Australian Research Council Discovery Early Career Researcher Award (Project Number: DE180100315) funded by the Australian Government. Anam Ahmad Khan, Joshua Newn and Namrata Srivastava are supported by scholarships under the Australian Commonwealth Government Research Training Program and Melbourne Research Scholarship.
\end{acks}

\bibliographystyle{ACM-Reference-Format}
\bibliography{sample-base}
\pagebreak 
\appendix

{

\section{Examples of Notes}

\begin{xltabular}{\textwidth}{p{2.4cm}p{6.1cm}p{4.6cm}}

\caption{Representative examples of the different kinds of notes made by participants in Study 2. The text passages are taken from research articles read by participants.} \\

\toprule 
\multicolumn{1}{l}{\textbf{Note Type}} & \multicolumn{1}{l}{\textbf{Reference Text Passages}} & \multicolumn{1}{l}{\textbf{Voice Note}} \\
\midrule 
\endfirsthead

\multicolumn{3}{c}
{\tablename\ \thetable{} -- \textit{continued from previous page}} \\
\toprule
\multicolumn{1}{l}{\textbf{Note Type}} & \multicolumn{1}{l}{\textbf{Reference Text Passages}} & \multicolumn{1}{l}{\textbf{Voice Note}} \\
\midrule 
\endhead

\\ \hline \multicolumn{3}{r}{{\textit{Continued on next page}}} \\
\endfoot

\bottomrule
\endlastfoot

\textbf{Short} & {\small For each attribute, CACC first finds the maximum and minimum of attributes and then forms a set of all values of each attribute in the ascending order. For all possible interval boundaries and all the midpoints of all the adjacent boundaries in the set are obtained and keeping the maximum CACC value and then partition this attribute accordingly into intervals. --- \cite{chaudhari2014}} & ``\textit{{\small Over here the process of CACC is explained that how it finds the maximum and minimum of each attributes and then later forms a set of values of each attribute which they can take}}.'' (P14)\\ \\

& {\small We used a qualitative approach based on 2 focus groups: one composed of 6 general practitioners and the other of 6 practice nurses. Discussion guidelines on the topics to be investigated were provided to the moderator. Sessions were audio-recorded and transcribed verbatim. Thematic analysis was performed using the ATLAS-ti software. --- \cite{Bermejo2019}} & ``\textit{{\small Okay. So they use the thematic analysis with Atlas ti software.}}'' (P5)\\ \\
      
& {\small  The learner experiences a variety of possible emotions, depending on the context, the amount of change, and whether important goals are blocked. However, this type of affective arousal that accompanies learning is still not well understood. For example, researchers have yet to narrow down the emotions that accompany deep level learning of conceptual material. The consequential impact of the emotions on knowledge acquisition and transfer is still not well understood. --- \cite{Mello2007}} & ``\textit{{\small okay so  here they talk about some of the questions that need further explanation by the research, when goals are blocked about then they may go to possible emotions}}.'' (P16)\\ \\

\midrule

\textbf{Reflective} & {\small Similar to the many approaches in previous work, we focus on opportunities for digital eyewear designs that do not rely on microdisplays or image generation, which also frees us from requiring traditional display driver circuitry with dedicated power and thermal implications. --- \cite{olwal2018} }. & ``\textit{{\small And so this subtopic about minimum near-eye display LEDs, I think that's what I was wondering about. I was thinking to propose a low-resolution eyewear, so it will have the low-resolution displays, but I wasn't sure if it can be done by transparent displays, or it should be projections.}}'' (P4) \\ \\ 
      
& {\small Results: The available operating systems and browsers and the lack of suitable spaces and time were reported as the main difficulties with the vCoP. The vCoP was perceived to be a flexible learning mode that provided up-todate resources applicable to routine practice and offered a space for the exchange of experiences and approaches. --- \cite{Bermejo2019} }. & ``\textit{{\small Okay. So my study basically goes against the results here so I should add this as a contribution to my work that we found that operating systems and browsers are okay. They are not a problem, not an issue because the system that we use was really good and with time use. I should add in my thesis as a barrier in the literature, then I would say that we fixed that barrier.}}'' (P5) \\ \\
      
& {\small Modern vehicles are increasingly becoming“smarter.” They are typically equipped with rich instrumentation such as GPS receivers, Internet access, wireless local area network (e.g, a dedicated short range communications technology), increasingly powerful electronic control units (ECUs), and engine sensors to periodically measure sub-system proper- --- \cite{ali2015} }. & ``\textit{{\small Reminds me about the vision paper i have read before, which i can cite as modern vehicles }}'' (P32) \\ \\

\textbf{Comprehensive} & {\small MIS is an integrated, user-machine system for providing information to support operations, management, analysis and decision-making functions in an organization. The system utilizes computer hardware and software; manual procedures; models for analysis, planning, control and decision making; and a database.}

\vspace{5pt}

{\small As the field of MIS has grown and developed, a clear contingency approach has emerged. The contingency approach suggests that a number of variables influence the performance of information systems; the better the "fit" between these variables and the design and use of the MIS, the better the MIS performance. Furthermore, there is an assumed "fit" between MIS performance and organizational performance; in many studies, as shown below, this "fit" is assumed rather than demonstrated. --- \cite{weill1989}} & ``\textit{{\small So to summarize, this section gives an overview of the MIS discipline so in the sense that the management of the information system is probably focuses on a system centric approach to managing information so that the different functions in the organizations can be managed. To what extent the information can be managed in the organization and then it starts with giving an overview on the performance. I am interested in the information system the goal is the performance of the contingency variable relates to the performance of the information system itself.}}'' (P7) \\ \\
     
& {\small The implementation of Healthcare Information systems, such as Computerized Physicians Order Entry (CPOE), Clinical Information Systems (CIS) or Electronic Medical Records (EMR) and, more recently, Hospital Information System (HIS), is supposed to have various benefits for the medical practices, as providing easy access to documentation of patients records and accurate them, billing management, reducing potential medical errors, and improving the quality of patient care.} 
      
\vspace{5pt}
     
{\small However, previous studies have shown the use of HIS has led to unintended consequences in the actual work practices, such as increased documentation time, incompatibility with clinical workflow, increasing more interruptions in medical work and system-introduced errors in patients care. --- \cite{Taddei2015}} & ``\textit{{\small The information system implementation in healthcare is beneficial because of improvement in health practices and also reduce potential medical errors  and also improve the quality of patient care. But, there's also the downside to installing such systems and the problems that arise due to certain decision permutation changes in work practices, too much time taken using the system to upload information in the system and also problems of the information systems not being compatible with the actual workflow in the hospital. Also, there could be some system introduced errors while taking care of patients.}}'' (P17) \\
     
& {\small Much research has been done in the area of thermal anomaly detection in data centers. Some approaches opted to use Linear Regression (LR) models not only to detect thermal anomalies, but also to rank and prioritize them. Some others, have enhanced thermal efficiency with Computational Fluid Dynamics (CFD) in order to rearrange racks and improve the cooling airflow. Reference [14] implements thermal cameras and correlation models to detect temperature deviations. Besides, they propose a novel thermal-anomaly aware allocation policy to reallocate incoming workload.}
      
\vspace{5pt}
     
{\small Further research works, more similar to ours, make use of Principal Component Analysis (PCA) to detect thermal anomalies in data centers. The use of ANNs is not new either, Yuan et al. [16] propose the implementing hierarchichal ANNs to detect a wide range of temperature anomalies. However, they need to evaluate several metrics such as CPU usage, temperatures of inlet and CPU and fan speed, increasing its complexity. --- \cite{aransay2015}} & ``\textit{{\small They showed a brief summary of the work, so the, based on this discussion I think there are couple of methods to detect thermal anomaly. So one is using a linear regressions, like simple models. The second one is complex models or, and of based in the final methods, which they have used up ANN models. The use of their work is not new, it has been proposed before using hierarchichal ANN. However some or several metrics such as CPU were missing}}'' (P26) \\ \\
\end{xltabular}
}

\end{document}